\documentclass[conference]{IEEEtran}
\IEEEoverridecommandlockouts
\usepackage{authblk}
\usepackage{cite}
\usepackage{amsmath,amssymb,amsfonts}
\usepackage{algorithmic}
\usepackage{graphicx}
\usepackage{textcomp}
\usepackage{xcolor}
\usepackage{float}
\usepackage{subcaption} 
\def\BibTeX{{\rm B\kern-.05em{\sc i\kern-.025em b}\kern-.08em
    T\kern-.1667em\lower.7ex\hbox{E}\kern-.125emX}}

\usepackage{pifont}
\usepackage{subfig}
\usepackage{colortbl}
\usepackage{braket}
\usepackage{comment}

\newcommand{\CellWithForceBreak}[2][c]{
\begin{tabular}[#1]{@{}c@{}}#2\end{tabular}}

\usepackage{balance}
\usepackage{xcolor}
\usepackage[ruled,vlined,noend]{algorithm2e}

\SetCommentSty{mycommfont}
\SetAlFnt{\small}
\SetAlCapFnt{\small}
\SetAlCapNameFnt{\small}

\usepackage{soul}
\def\BibTeX{{\rm B\kern-.05em{\sc i\kern-.025em b}\kern-.08em
    T\kern-.1667em\lower.7ex\hbox{E}\kern-.125emX}}

\begin{document}

\title{OPAQUE: Obfuscating Phase in Quantum Circuit Compilation for Efficient IP Protection }

\author[1]{Anees Rehman}
\author[2]{Vincent Langford}
\author[2*]{Yuntao Liu}
\affil[1]{Independent Researcher, Pakistan }
\affil[2]{Department of Electrical and Computer Engineering, Lehigh University, PA, USA}
\affil[*]{yule24@lehigh.edu}

\maketitle

\begin{abstract}
Quantum compilers play a crucial role in quantum computing by converting these algorithmic quantum circuits into forms compatible with specific quantum computer hardware. However, untrusted quantum compilers present considerable risks, including the potential theft of quantum circuit intellectual property (IP) and compromise of the functionality (e.g. Trojan insertion).
Quantum circuit obfuscation techniques protect quantum IP by transforming a quantum circuit into a key-dependent version before compilation and restoring the compiled circuit's functionality with the correct key. This prevents the untrusted compiler from knowing the circuit's original functionality. Existing quantum circuit obfuscation techniques focus on inserting key qubits to control key gates. One added key gate can represent at most one Boolean key bit.
In this paper, we propose OPAQUE, a phase-based quantum circuit obfuscation approach where we use the angle of rotation gates as the secret keys. The rotation angle is a continuous value, which makes it possible to represent multiple key bits. Moreover, phase gates are usually implemented as virtual gates in quantum hardware, diminishing their cost and impact on accuracy.
\end{abstract}

\section{Introduction}
Quantum computing has been growing rapidly and has demonstrated the capability to efficiently solve some complex problems that can take the fastest classical computer an astronomical time to solve \cite{blogMeetWillow}. Central to every quantum computer are qubits, which, based on quantum mechanics principles, differ from classical bits. While classical bits represent either 0 or 1, qubits can be in \textit{superposition} of $\ket 0$ and $\ket{1}$. Quantum gates control one or multiple qubits in a similar fashion to how logic gates control classical bits. Single qubit quantum gates include Hadamard, Pauli, and phase gates, etc. Multi-qubit gates creates \textit{entanglement} among two or more qubits. Such gates include controlled Pauli, swap, and Toffoli gates, etc. Both single and multi-qubit gates are crucial for implementing intricate quantum algorithms. Interconnecting these gates forms quantum circuits, bridging the gap between quantum algorithms and hardware (quantum processors). Presently, several cloud providers, such as IBM Quantum \cite{chow2021ibm}, Amazon Braket \cite{gonzalez2021cloud}, and Microsoft Azure \cite{prateek2023quantum}, offer access to quantum computing resources. To execute a quantum algorithm, users send their designs to a quantum compiler, like Qiskit \cite{qiskit2024}, which optimizes and schedules circuits for designated quantum hardware.
Despite significant strides in quantum computing, it is still resource-demanding and time consuming to reach optimal performance when designing quantum algorithms. Consequently, quantum circuit designs are deemed essential intellectual property \cite{aboy2022mapping}. The compilation stage translates the algorithmic (a.k.a. idealized) quantum circuit into a functionally equivalent form that is executable by the target quantum computer. In this process, all quantum gates are mapped to those supported by the quantum computer and swap gates are inserted to ensure that physical qubit entanglement facilitates the execution of multi-qubit operations in the circuit. Additionally, various optimizations can be applied to enhance the efficiency of compiled circuits. There are inherent security risks in this process, as untrusted compilers could potentially exploit and abuse the quantum designs, potentially leading to the counterfeiting of quantum designs \cite{yang2024multi} or the insertion of trojans \cite{das2023trojannet, roy2024hardware} . Given the high value of these quantum circuits as intellectual property, safeguarding them against compiler-based attacks is of paramount importance.

In this paper, we propose OPAQUE, a novel method to obfuscate quantum circuits where phase gates are inserted at selected locations for obfuscation and their rotation angles serve as key values. One key advantage of this approach is that the key value is continuous. Compared to existing approaches which use Boolean keys \cite{topaloglu2023quantum}, our approach capitalizes on the continuous nature of quantum circuits, improving the efficiency of key insertion and hence the security levels. 
Our paper has the following contribution.
\begin{itemize}
   \item We propose OPAQUE, a novel phase-based obfuscation for quantum circuits against untrusted compilers. 
   \item We introduce a process to determine obfuscation location based on a topological analysis of the quantum circuit.
   \item We present a security analysis that highlights OPAQUE's advantages in incorporating key bits more efficiently than previous approaches thus reaching higher security levels.
   \item Experimental results demonstrate the high efficacy and low overhead of OPAQUE.
\end{itemize}

\section{Background}

\subsection{The Phase of a Qubit} \label{ssec:phase}
\begin{figure}[h]
    \captionsetup{font=small} 
    \centering
    \includegraphics[width=0.2\textwidth, trim={1 1 1 1}, clip]{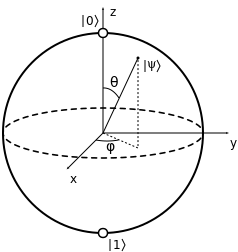}
    \caption{The Bloch sphere.}
    \label{fig:bloch_sphere}
\end{figure}

The fundamental unit of quantum computing, the quantum bit or qubit, is analogous in concept to the classical Boolean bit. A qubit is characterized by two basis states, represented in bracket notation as $\ket 0$ and $\ket 1$. These basis states can be viewed as two-dimensional vectors where $\ket 0 = [1, 0]^T$ and $\ket 1 = [0, 1]^T$. A qubit state $\ket \psi$ may be expressed as $\ket \psi = \alpha \ket 0 + \beta \ket 1 = [\alpha, \beta]^T$ where $\alpha$ and $\beta$ are complex numbers and $|\alpha|^2+|\beta|^2=1$. The state space of a qubit can be geometrically represented with a Bloch sphere, shown in Figure \ref{fig:bloch_sphere}. The poles of the Bloch sphere represent the base states, with $\ket 0$ at the north pole and $\ket 1$ at the south pole. Each location on the sphere represents a quantum superposition of the base states.
There are two angles $\theta$ and $\phi$ defined in the Bloch sphere, with $\theta$ denoting the qubit's magnitude and $\phi$ the relative phase of the two basis states. 
Using $\theta$ and $\phi$ , $\alpha$ and $\beta$ can be expressed as $\alpha =\cos \frac{\theta}{2}$, $\beta = e^{i\varphi}\sin \frac{\theta}{2}$.
Notice that the relative phase is not the global phase of the qubit state. A qubit state is mapped to the same point on the Bloch sphere regardless of its global phase. In this paper, we use the relative phase of a qubit for obfuscation rather than the global phase.

\subsection{Phase Gates}
In quantum computing, rotation gates are fundamental elements that allow us to manipulate the state of qubits. These gates perform rotations around specific axes on the Bloch sphere. One of the commonly used rotation gates is the $R_Z$ gate which rotates a qubit around the $Z$-axis. The $R_Z$ gate is parametrized by an angle $\delta$ and its operation can be represented as $R_Z(\delta) = \begin{pmatrix} e^{-i\delta/2} & 0 \\ 0 & e^{i\delta/2} \end{pmatrix}$. This gate is particularly useful in quantum algorithms as it allows for phase shifts without changing the probability amplitudes of the computational basis states. Although the $R_Z$ gate does not change a qubit's probability amplitudes by itself, it can have an impact when used in conjunction with other gates. For example, let us consider the case where an $R_Z$ gate is sandwiched between two Hadamard gates. When the rotation angle is zero, the three concatenated gates $HR_Z(0)H$ equals an identity gate $I$. If the phase shift in the $R_Z$ gate is $\frac{\pi}{2}$, then $HR_Z(\frac{\pi}{2})H$ is equivalent to a NOT gate $X$. We will use $R_Z$ gates as the key gates in OPAQUE.

\subsection{Controlled Quantum Gates and Quantum Entanglement}
Controlled quantum gates operate on two or more qubits and create entanglement among them. In other words, the state of each qubit after these gates are dependent on both qubits' input states. 
Common controlled quantum gates include $C_X$ (controlled-X or controlled-NOT), $C_Y$ (controlled-Y), and $C_Z$ (controlled-Z) gates. In a classical view, the two qubits in these gates are labeled as the ``control'' qubit and the ``target'' qubit, respectively. However, this does not reflect the two-way nature of quantum entanglement. In fact, the ``target'' qubit also affects the state of the ``control'' qubit unless the ``control'' qubit is in the basis state $\ket 0$ or $\ket 1$ (i.e. not in superposition). One famous example of this fact is the phase kickback effect where the ``target'' qubit can change the phase of the ``control'' qubit \cite{cleve1998quantum}, which has widespread usage in quantum algorithms such as quantum Fourier transform, quantum phase estimation, and Grover's algorithm \cite{chapeau2020fourier, long2001grover}. 
In existing quantum obfuscation techniques such as \cite{topaloglu2023quantum} and \cite{liu2025e}, $C_X$ gates are often used to incorporate key values from the key (ancilla) qubit(s) which are considered to be either $\ket 0$ or $\ket 1$. This underutilizes both the qubit state space for keys and the ability of controlled qubits to spread error both ways.

\subsection{Quantum Circuit Compilers}
Quantum circuit compilers converts idealized quantum circuit into a form compatible for execution on the target quantum hardware. Each target quantum hardware has a set of supported gates and constraints on which qubits can be entangled. The compiler must tailor the compiled circuit to the hardware-specific constraints. This process involves mapping the circuit qubits to the hardware qubits. In case there is no subset of hardware qubits that can support the entanglement required by the quantum circuit, multiple qubits in the hardware must be ``pooled'' to support the entanglement topology.

\section{Threat Models and Existing Defenses}
\begin{figure}[htb]
\captionsetup{font=small} 
    \centering
    \includegraphics[width=0.5 \textwidth]{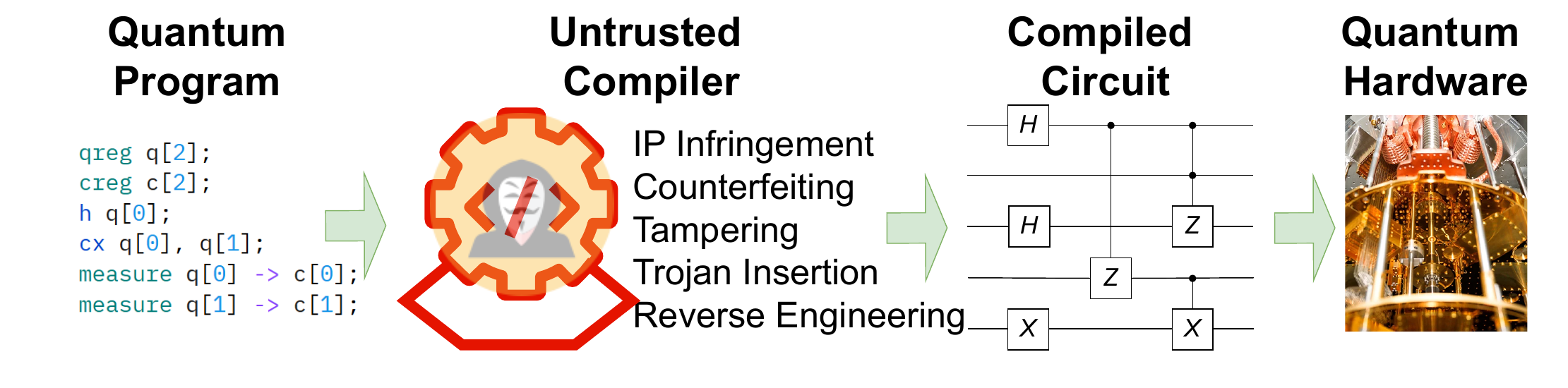}
    \caption{The considered threat model of untrusted quantum circuit compilers.}
    \label{fig:threat_model}
\end{figure}

This paper examines the potential threats posed by third-party quantum circuit compilers. These compilers offer numerous benefits, including optimizing quantum circuits for various quantum computing platforms and implementing error reduction techniques \cite{smith2020open,salm2021automating}. Qulic \cite{smith2020open} and TKET \cite{sivarajah2020t},  \cite{qiskit2024} and Cirq \cite{hancockcirq} are some of the most notable quantum circuit compilers. These compilers are mentioned to highlight the widespread use of third-party quantum compilers, without suggesting that they have any malicious intent. Some quantum circuit compilers (e.g. Qiskit and Circ) are native to quantum computers from one vendor but also support other platforms, functioning as third-party compilers. Despite the benefits of third-party compilers, they pose significant risks to the intellectual property (IP) of quantum circuits. Using a compiler necessitates unveiling the entire circuit, which could lead to IP theft. Additionally, compilers may alter the circuit's design without permission, potentially incorporating malicious functionality. Figure \ref{fig:threat_model} illustrates these potential risks. The threat model presented here is consistent with those used in previous quantum adversary studies \cite{das2023randomized, suresh2021short, saki2021split, topaloglu2023quantum, liu2025e}.

\subsection{Existing Quantum Circuit Obfuscation Techniques}

Researchers have investigated various methods to protect quantum circuits from unauthorized access by third-party compilers. Strategies include inserting random reversible circuits \cite{das2023randomized, suresh2021short, das2024secure}, using ancilla qubits to represent keys to lock quantum circuits \cite{topaloglu2023quantum, liu2025e}, and dividing the circuit for separate compilation \cite{saki2021split}. In works \cite{das2023randomized} and \cite{das2024secure}, random reversible circuits are generated and inserted at the beginning, middle, or end of the original quantum circuit for obfuscation. The original circuit's functionality is restored after compilation by reversing these additions in the same locations. A similar technique is used in \cite{suresh2021short}. 
These type of methods bear significant overhead, because extra quantum gates must be inserted on all qubits for obfuscation. This will increase both latency and error of the quantum circuit.
Split compilation techniques such as \cite{saki2021split} may still suffer from a brute-force attack when the same compiler compiles two adjacent parts of the circuit or two colluding compilers are used. 
\cite{topaloglu2023quantum} and \cite{liu2025e} proposed methods similar to traditional logic locking, utilizing additional key qubits and key gates to obfuscate the original quantum circuit functionality and topology. While these approaches can be effective, they are also very resource consuming. The locking approach proposed in \cite{topaloglu2023quantum} needs a separate qubit for each (binary) key bit. Although the enhanced locking method in \cite{liu2025e} addressed this problem by condensing all key bits onto a single added qubit, it still needs a Hadamard gate and a controlled gate to deliver each key bit into the quantum circuit. These added gates accumulate noise and degrade the circuit's accuracy. This also means that, under qubit and gate budget constraints, the security level that can be achieved using existing quantum circuit locking techniques can be rather limited.

Other research directions related to protecting quantum circuit IP include quantum trusted execution environments on untrusted clouds \cite{trochatos2023hardware, trochatos2024dynamic} and thwarting side-channel and fault-injection attacks on quantum hardware \cite{lu2024quantum, maurya2024understanding, bell2022reconstructing, xu2023exploration, erata2024quantum}. These research directions are also very important but outside the scope of our paper.

\begin{figure*}
    \centering
    \includegraphics[width=\textwidth]{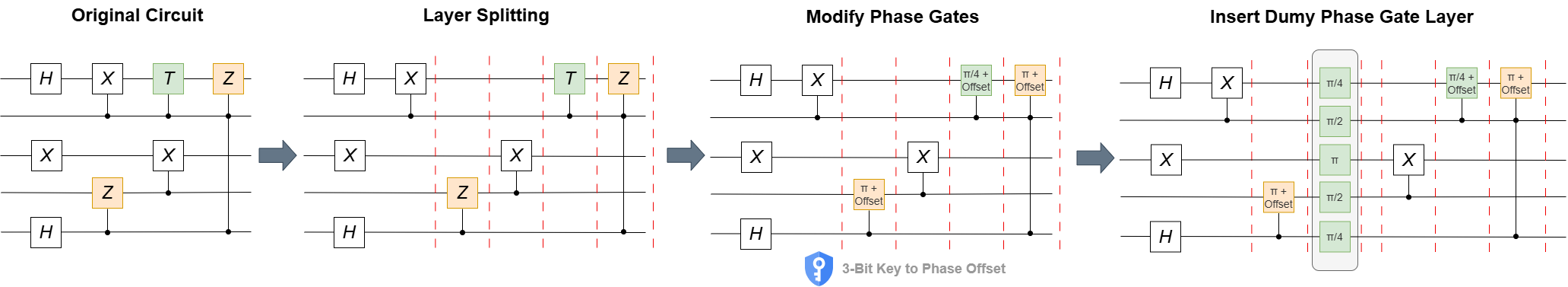}
    \caption{Illustration of the OPAQUE phase obfuscation technique. The quantum circuit is split into layers that contain only phase gates and non-phase gates. Barriers are added after phase gate layers to ensure that the location is retained after compilation. All the phase angles will be obfuscated according to the obfuscation key and random layer of phase gates will be added with random phase angles.}
    \label{fig:phase_obfuscation}
\end{figure*}

\section{Phase Obfuscation for Quantum Circuits}
The aim of the OPAQUE framework is to introduce a more efficient quantum circuit locking technique where multiple Boolean key bits can be represented by one quantum gate. This can be achieved by using $R_Z$ phase gates to obfuscate the functionality of quantum circuits and the rotation angle is the secret key. Quantum circuit designers can obfuscate the circuits with $R_Z$ gates of randomized angles for obfuscation before compilation. After the circuit has been compiled, the rotation angles in the phase gates will be replaced with the correct angle for de-obfuscation.

\subsection{Obfuscation Procedure}
Choosing the obfuscation location and determining the rotation angles are the two primary tasks in the phase obfuscation process. An ideal obfuscation location should cause maximum error at the output qubits of the circuit. In classical circuit obfuscation, key gates are usually located at nodes whose transitive fan-out cone covers the most output bits \cite{yasin2015improving}. Similarly, for quantum circuit obfuscation, it is desirable to find locations with high output error impact to insert key gates.
One way to guarantee a maximum fan-out cone is to insert an entire layer of phase gates. To this end, we convert the quantum circuit into a directed acyclic graph (DAG) and organize quantum gates into layers and make sure that each layer only has non-phase gates or phase gates (not a combination of them). We also insert layers of dummy phase gates depending on the budget. The rotation angles of the added phase gates are derived using the key value.
The detailed process is given in Algorithm \ref{alg:lock} and illustrated in Figure \ref{fig:phase_obfuscation}.








\begin{algorithm}[htb]
\DontPrintSemicolon
\SetAlgoLined
\SetNoFillComment
\LinesNotNumbered 
\caption{Quantum Circuit Obfuscation and Phase Locking}\label{alg:lock}

\KwData{\textbf{C}: Quantum circuit}
\KwIn{Quantum circuit $Q$, secret key $K$}
\KwOut{Obfuscated quantum circuit $Q'$, stored phase gate information $\mathcal{P}$}

\tcc{Step 1: Extract Layers}
Convert circuit to DAG representation and extract layers.\;

\tcc{Step 2: Layer Splitting}
Split each layer into:
\begin{itemize}
    \item Phase gate layer (e.g., $Rz, P, S, T$)
    \item Non-phase gate layer (e.g., $X, Y, Z, H, CX$)
    \item Enclose them within barriers for isolation.
\end{itemize}

\tcc{Step 3: Store Phase Gate Information}
Store phase gate types and their respective phases as $\mathcal{P}$.\;

\tcc{Step 4: Modify Phase Gates}
Modify each phase gate by adding a random phase shift derived from the secret key $K$.\;

\tcc{Step 5: Insert Random Phase Gate Layers}
Select random positions in the circuit and:
\begin{itemize}
    \item Insert dummy layers of randomly generated phase gates.
    \item Enclose them within barriers for isolation.
\end{itemize}

\tcc{Step 6: Convert Back to Standard Representation}
Transform the recovered DAG representation back into a quantum circuit format.\;

\KwRet{Obfuscated quantum circuit $Q'$, stored phase gate information $\mathcal{P}$}
\end{algorithm}

\subsection{De-obfuscation Procedure}

After the compilation process is completed, the correct key and phase gates information is needed to recover the original circuit. 
Specifically, the phase angle in added phase gates will be reversed using the de-obfuscation key. If this key matches the locking key, the correct phase angles will be restored, and the de-obfuscated compiled quantum circuit will have correct functionality when executed on quantum computers. If an incorrect de-obfuscation key is used, OPAQUE ensures that the resulting circuit will experience significant functional corruption, as we will describe in our experiment results.
The detailed process is given in Algorithm \ref{alg:unlock}.

\begin{algorithm}[htb]
\DontPrintSemicolon
\SetAlgoLined
\SetNoFillComment
\LinesNotNumbered 
\caption{Quantum Circuit Deobfuscation and Phase Recovery}\label{alg:unlock}

\KwData{\textbf{C'}: obfuscated quantum circuit, stored phase gate information $\mathcal{P}$}
\KwIn{Obfuscated quantum circuit $Q'$, stored phase gate information $\mathcal{P}$, secret key $K$}
\KwOut{Recovered quantum circuit $Q$}

\tcc{Step 1: Extract Layers}
Convert circuit to DAG representation and extract layers.\;

\tcc{Step 2: Identify and Remove Random Phase Gate Layers}
Use stored phase gate information $\mathcal{P}$ to identify dummy random phase gate layers and remove the randomly inserted dummy phase gate layers enclosed within barriers.\;

\tcc{Step 3: Reverse Phase Modifications}
For each modified phase gate, subtract the random phase shift derived from $K$ to restore its original phase.\;

\tcc{Step 4: Reconstruct Circuit Layers}
Merge phase gate layers with their corresponding non-phase gate layers and remove barriers.\;

\tcc{Step 5: Convert Back to Standard Representation}
Transform the recovered DAG representation back into a quantum circuit format.\;

\KwRet{Recovered quantum circuit $Q$}
\end{algorithm}



\subsection{Improvements over Prior Work}
One key disadvantage of the existing locking-based quantum circuit obfuscation approaches is that they either need a separate qubit \cite{topaloglu2023quantum} or a controlled quantum gate \cite{liu2025e} to incorporate a Boolean key bit. In OPAQUE, as the rotation angle is a continuous value between $0$ and $2\pi$, it can theoretically incorporate any number of Boolean key bits. However, not all angles are equally likely to be the correct angle. For example, the $S$ and $T$ gates changes the phase by $\frac{\pi}{2}$ and $\frac{\pi}{4}$, respectively. They belong to the Clifford group \cite{tolar2018clifford}, a set of gates widely used in quantum algorithms such as quantum error correction, entanglement distillation, and randomized benchmarking. A combination of these gates can implement any phase change by multiples of $\frac{\pi}{4}$. Hence, these angles are more likely to be the correct angles, which means an adversary has a better chance guessing these angles. There are 8 multiples of $\frac{\pi}{4}$ within $[0,2\pi)$. Therefore, each phase gate can encode 3 Boolean key bits. Notice that this is an underestimation, because the actual phase change angles are not limited to multiples of $\frac{\pi}{4}$.

Phase gates can be implemented as virtual gates in quantum hardware. Instead of using a separate gate for phase change, the hardware can simply adjust the phase for the following operation \cite{vezvaee2024virtual}. This gives OPAQUE three additional advantages: speed, fidelity, and flexibility. Since virtual gates are not physical gates, they do not introduce delay or noise into the quantum circuit. In contrast, the key gates introduced by existing quantum circuit obfuscation approaches are all physical gates, which increases the delay and noise in the obfuscated circuits. In order not to incur additional delay, the quantum gates for obfuscation can be inserted in \textit{idle qubit windows} (i.e. on idle durations of a qubits when other qubits are busy) \cite{rasmussen2024time}. This saves time at the cost of fewer choices for obfuscations. Since virtual phase gates do not cause any additional delay, they can be placed anywhere without impacting the timing of the circuit.

\section{Experiments and Evaluation}
\subsection{Experimental Setup}

We selected benchmark circuits from \textit{RevLib} \cite{wille2008revlib} and \textit{QASMBench} \cite{li2023qasmbench}, resources that have been extensively used in earlier research on quantum circuit compilation. These benchmark circuits feature a range of gate operations. Our experimental analysis was carried out employing the IBM Qiskit framework to compile and simulate quantum circuits. To mimic realistic simulation conditions, we used the \textit{FakeValencia} backend from Qiskit \cite{qiskit2024}, reflecting the noise model of the real \textit{ibmq-valencia} device. Simulations were executed with 1,000 shots to ensure results with statistical significance. We simulated both the original and obfuscated circuits with the same backend, which means any observed differences can be attributed to the method used rather than shifts in the simulation environment.

\subsection{Metrics for Evaluation}

We use \textit{Total Variation Distance (TVD)} as the metric for obfuscation quality. In statistics, TVD is a (dis)similarity metric between two probability distributions. It calculates the total absolute differences between the probabilities of each event in the two distributions. 
Given the inherent probabilistic nature of quantum circuit output measurements, the TVD metric is suited to measure the difference between the output distributions of the original and obfuscated circuits. 
The difference in output distribution reflects the obfuscation quality in terms of corrupting the correct output. Considering a quantum circuit with $b$ output qubits and measured with $N$ shots, the TVD forumla is given in Equation \ref{eq:tvd}.
\begin{equation}
    TVD = \frac{\sum_{i=0}^{2^b-1} |y_{i,orig} - y_{i,alter}|}{2N}
\label{eq:tvd}
\end{equation}
Notice that there are $2^b$ possible output types. We use $y_{i,orig}$ and $y_{i,alter}$ to represent the number of times that the original and altered quantum circuits output value $i$, respectively. 


\begin{table*}[h]
\captionsetup{font=small} 
\centering
\begin{tabular}{|c|c|c|c|c|c|c|c|c|c|c|c|c|}
\hline
\textbf{\CellWithForceBreak{Name \\ }} & 
\textbf{\CellWithForceBreak{\# \\ Qubits}} &
\textbf{\CellWithForceBreak{Depth \\ Orig.}} & 
\textbf{\CellWithForceBreak{Depth \\ Obf.}} &
\textbf{\CellWithForceBreak{\# phase \\ gates \\ Orig.}} &
\textbf{\CellWithForceBreak{\# dummy \\ phase \\ gates}} &
\textbf{\CellWithForceBreak{\# \\ equiv. \\ key bits}} &
\textbf{\CellWithForceBreak{TVD \\ Obf.}} &
\textbf{\CellWithForceBreak{TVD \\ Deobf.}} &
\textbf{\CellWithForceBreak{TVD \\ Orig.}} &
\textbf{\CellWithForceBreak{TVD \\ Loss}} \\ \hline
\text{adder\_n4}          &4 &12   &17    &9   &16 &75    &0.50  &0.21 &0.15 &0.06 \\ \hline
\text{basis\_trotter\_n4} &4 &815  &1088  &492 &16 &1524  &0.19  &0.12 &0.09 &0.03 \\ \hline
\text{fredkin\_n3}        &3 &12   &19    &7   &12 &57    &0.66  &0.35 &0.27 &0.08  \\ \hline
\text{basis\_change\_n3}  &3 &22   &27    &0   &12 &36    &0.49  &0.24 &0.19 &0.05 \\ \hline
\text{wstate\_n3}         &3 &6    &11    &0   &12 &36    &0.67  &0.40 &0.31 &0.09 \\ \hline
\end{tabular}
\caption{\textit{OpenQASMBench} benchmark information, security parameters used in our experiments, and fidelity change data.}
\label{tab:circuit_parameters}
\end{table*}

\subsection{Result Analysis}
\begin{figure}[!htp]
    \centering
    \captionsetup{font=small} 
    \includegraphics[width=0.48\textwidth]{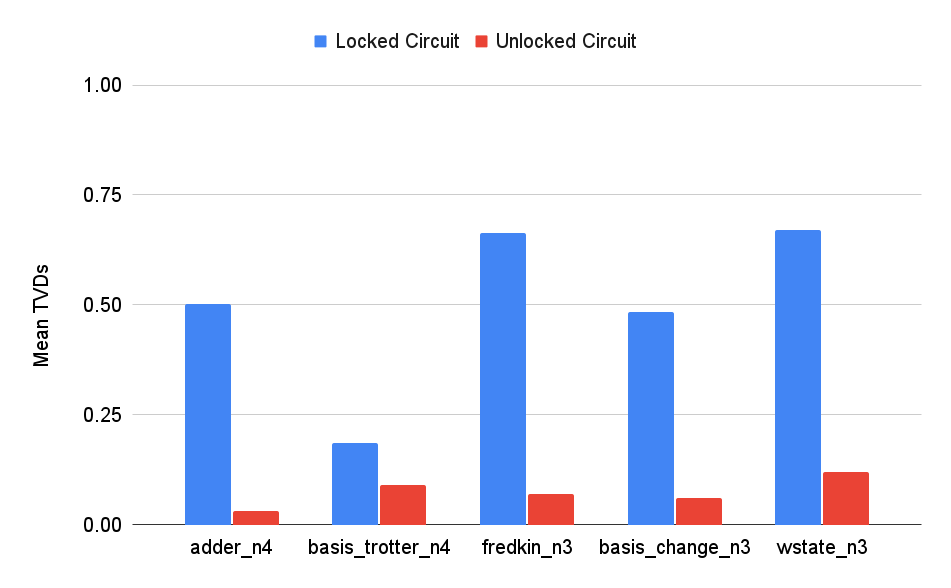}
    \caption{Distribution of Total Variation Distance (TVD) of benchmark circuits from the OpenQASM benchmark suite: TVD of obfuscated circuit and restored circuit are calculated and shown respectively. Selected circuits are simulated using Qiskit and AerSimulator backend.}
    \label{fig:result_vd}
\end{figure}

In this section, we present our experimental results, starting with the outcomes from the \textit{RevLib} and \textit{OpenQASMBench} benchmarks simulated using the noise-enabled \textit{AerSimulator} backend in Qiskit. 
Figure \ref{fig:result_vd} illustrates a comparison of mean TVD values between simulated and theoretical output values across different circuit benchmarks in \textit{OpenQASMBench} for the obfuscated case and the restored case. The TVD of the restored circuits w.r.t. the theoretical output should be very low as they are caused only by the noise of inherently in the quantum circuits. The TVD of the obfuscated should be much larger, as it is the result of obfuscation. The TVD of the unlocked circuit shows the intrinsic noise in quantum computing.
We do not report any data for the \textit{RevLib} benchmarks here. This is because \textit{RevLib} benchmarks merely use quantum gates to emulate Boolean expressions. In our experiments, each qubit is initialized as $\ket 0$. Hence, there is no quantum state superposition in these benchmarks and the state of any qubit a any time is either $\ket 0$ or $\ket 1$, making any phase change ineffective. Hence, OPAQUE is not suitable for obfuscating this type of circuits under this scenario. This should not be considered a weakness of OPAQUE, as such problems can be solved efficiently by classical computers.
In out future work, we will consider \textit{RevLib} benchmarks with superpositioned input qubit states and investigate the effectiveness of OPAQUE under this scenario. 

\subsection{Security and Overhead Analysis}
In Table \ref{tab:circuit_parameters}, we list the security and cost metrics of OPAQUE applied on the \textit{OpenQASMBench} benchmarks. Specifically, we use the number of key bits to quantify the security level, the increase in circuit depth and gate count to quantify the cost, and the fidelity difference between the deobfuscated circuit and the original circuit to quantify the performance overhead.

\subsubsection{Number of Key Bits}
The efficiency to incorporate key bits is a unique advantage of OPAQUE. Specifically, key bits are incorporated into both existing and added phase gates at a rate of 3 bits per phase gate. In our benchmarks, depending on how many existing phase gates are present and how many we can add in added dummy phase gate layers, we have incorporated between 36 and 1524 key bits in each benchmark. There are unprecedented number of key bit counts for quantum circuit obfuscation.

\subsubsection{Circuit Depth}
We can observe some increase in circuit depth due to the OPAQUE obfuscation process. This is because we re-organize the circuits and put phase gates in dedicated layers, which ensures that barriers can be added after phase gate layers in order to track the phase gate insertion locations. By doing so, the depth of the circuit has increased, and the increase is more significant in circuits where there are more phase gates in the original circuit.

\subsubsection{Fidelity of Unlocked Circuits}
The fidelity decrease measured by the TVD loss between the original and the deobfuscated circuits are higher than Boolean obfuscation techniques on quantum circuits such as \cite{liu2025e}. This may be an inherent property of phase changes in quantum circuits as it may introduce more noise into the quantum circuit than flipping between the basis states $\ket 0$ and $\ket 1$. However, the number of key bits incorporated in the same number of key bits is at least 3x larger in OPAQUE, so the fidelity loss per key bit is not essentially higher in OPAQUE.

\section{Conclusion}
In this paper, we introduce OPAQUE, a novel phase-based obfuscation to thwart untrusted quantum circuit compiler attacks. Phase is a unique property of qubits and is not present in Boolean bits in classical computing. While existing untrusted compiler countermeasures generally use techniques migrated from classical digital circuits, OPAQUE is the first intrinsically quantum obfuscation methodology. OPAQUE takes advantage of the continuous nature of the phase and incorporate multiple key bits into one key gate, significantly improving the security level. Our experiments show that the obfuscated circuits have much greater TVD values compared to deobfuscated circuits, signifying the effectiveness of OPAQUE obfuscation. One limitation of OPAQUE is that it does not obfuscate the structure (entanglement topology) of the quantum circuit. In our future work, we will improve this technique towards both functional and structural obfuscation of quantum circuits. Additionally, we will consider the full input space of the quantum circuits, as well as addressing improving the cost and fidelity of phase obfuscation.

\balance


\bibliographystyle{IEEEtran.bst}
\bibliography{qref}
\end{document}